\documentclass[a4paper,aps,pre,twocolumn,showpacs,showkeys,preprintnumbers,amsmath,amssymb]{revtex4}
\usepackage[english]{babel}
\newcommand{\eq}[1]{eq. (\ref{#1})}
\newcommand{\beq}{\begin{equation}}
\newcommand{\eeq}{\end{equation}}
\newcommand{\beqa}{\begin{eqnarray}}
\newcommand{\eeqa}{\end{eqnarray}}

\newcommand{\vv}{{\mathbf v}}
\newcommand{\ee}{{\mathbf e}}
\newcommand{\m}{{\mathbf m}}
\newcommand{\x}{{\mathbf x}}
\newcommand{\z}{{\mathbf z}}
\newcommand{\h}{{\mathbf h}}
\newcommand{\0}{{\mathbf 0}}
\newcommand{\1}{{\mathbf 1}}
\newcommand{\J}{{\mathbf J}}
\newcommand{\U}{{\mathbf U}}
\newcommand{\T}{{\mathbf T}}
\newcommand{\M}{{\mathbf M}}
\newcommand{\e}{{\mathbf e}}
\newcommand{\E}{{\mathbf E}}
\newcommand{\C}{{\mathbf C}}
\newcommand{\G}{\Gamma}
\renewcommand{\S}{{\mathbf S}}
\newcommand{\g}{{\gamma}}
\renewcommand{\d}{\text{d}}
\newcommand{\tr}{\mbox{Tr}}
\renewcommand{\t}[1]{#1^{\top}}
\renewcommand{\(}{\left(}
\renewcommand{\)}{\right)}
\newcommand{\ave}[1]{\left\langle #1 \right\rangle}
\newcommand{\D}[1]{\frac{\d #1}{2\pi i #1}}

\begin{document}

\preprint{LU TP 03-18}

\title{Properties of Random Graphs with Hidden Color}

\author{Bo S\"oderberg}
\email{Bo.Soderberg@thep.lu.se}
\affiliation{Complex Systems Division, Dept. of Theoretical Physics, Lund University, Sweden}
\date{\today}

\begin{abstract}
We investigate in some detail a recently suggested general class of
ensembles of sparse undirected random graphs based on a hidden
stub-coloring, with or without the restriction to nondegenerate
graphs.  The calculability of local and global structural properties
of graphs from the resulting ensembles is demonstrated. Cluster size
statistics are derived with generating function techniques, yielding a
well-defined percolation threshold.  Explicit rules are derived for
the enumeration of small subgraphs. Duality and redundancy is
discussed, and subclasses corresponding to commonly studied models are
identified.
\end{abstract}



\maketitle

\section{Introduction}
\label{Intro}

Numerous phenomena in physics, molecular biology, social sciences and
information technology can be described in terms of networks, where
the nodes represent elementary units such as spins, genes, proteins,
people, or computers, while the links describe their interaction
structure.  The formation process of these networks typically is not
entirely deterministic, but involves stochastic components, and the
resulting networks can be viewed as {\em random graphs} -- random
members of a {\em statistical ensemble} of graphs.

We are primarily interested in {\em truly random} graphs, without any
prior distinction between individual nodes or groups of nodes, such as
an underlying lattice or other regular structure.
An example is the {\em classic model} of Erd\H{o}s and R\'enyi
\cite{ErRe60}, with a single parameter (in addition to the order $N$
of the graph) in the form of a real number $c$, such that each
possible edge is independently and randomly realized with a
probability $p=c/N$ (in the sparse version). The classic model has
been thoroughly studied in various versions, static as well as
evolving \cite{Boll01,Jans00,FKP89}.
Its asymptotic ($N\to \infty$) degree distribution is Poissonian with
average $c$, and it displays a {\em phase transition} in the form of a
percolation threshold at $c=1$, above which a giant component emerges
containing a finite fraction of the nodes in the thermodynamic limit
of large $N$.  For a long time this and related models dominated the
stage; however, they fail to describe the properties of most
real-world networks.

In the last decades a multitude of alternative random graph models
have been investigated, falling into two major categories. In a {\em
static model} a statistical ensemble of random graphs is considered
without bothering about how the graphs were formed
\cite{BenCan,MoRe95,Newm01,Sod02,BerLae02,Newm03}. A {\em dynamical
model} attempts to describe the random growth and evolution of a
network, leading to an evolving ensemble of graphs
\cite{AlBa00,Call01,DoMeSa01,Turo02}.

Here we will focus on {\em static descriptions} of random graphs in
terms of fixed statistical ensembles, bearing in mind that the
dynamics of real-world networks is not always directly observable, and
the comparison of model and reality typically has to be done based on
static properties as observed in snapshots of real networks.

For the inference of a particular model based on the observed
properties of real networks to be meaningful, a sufficiently general
formalism is desirable, where more specific models appear as special
cases of one and the same general class of graph ensembles.

In a recent paper \cite{CDRG}, a promising candidate for such a
general formalism was proposed; it will be referred to as {\em CDRG}
(for Colored Degree-based Random Graphs). It is based on a {\em hidden
coloring} of {\em stubs} (incidence points of edges upon a vertex) and
a specification of the colored stub distribution of vertices as well
as edges. This approach admits a unifying formalism for models of
symmetric, truly random graphs that are {\em sparse} (typical degrees
are finite and do not grow with the graph size $N$).

The resulting class of random graph ensembles
incorporates several commonly studied models, such as the classic
random graph, and random graphs with a given degree distribution
\cite{BenCan,Lucz92,MoRe98,Newm01}, as well as vertex-colored
extensions of these \cite{Sod02,Newm03}. Models with degree-biased
edge distributions \cite{BerLae02} also fit into this approach.
Furthermore, although the approach in its present form is restricted
to symmetric graphs, it has a natural extension to directed graphs,
which will be explored in forthcoming work.

The discussion in ref. \cite{CDRG} was restricted to ensembles of {\em
simple} (nondegenerate) graphs containing no cycles of length one
(self-couplings, or tadpoles) or two (double edges), based on the
restriction to simple graphs of an underlying
ensemble of {\em multigraphs} where such degeneracies are allowed.
Multigraph ensembles are interesting in their own right, and more
convenient for analytical purposes. Here, we will consider both types
of CDRG ensembles, denoting by {\em CDRG-s} the restriction to the
class of ensembles of simple graphs, and by {\em CDRG-m} the
unrestricted class of multigraph ensembles.
For generic ensembles of both types, we will present a theoretical
analysis of the properties of the resulting graphs, with an emphasis
on the analysis of observable local and global graph characteristics.

The computability of structural properties is an important factor for
the possibility of devising a systematic model inference scheme based
on the observed properties of real-world networks.  Both types of
ensemble admit an analysis of both global and local structural
properties of the resulting random graphs. The global connectivity
properties of a graph can be analyzed in terms of the size
distribution of connected components, for which a generating function
analysis was devised in ref. \cite{CDRG}. Local structural properties
are associated with the frequencies of appearance of small subgraphs;
also these will be shown to be asymptotically computable in both types of
ensemble.

The remainder of this article has the following structure.
In section \ref{Notation}, we will define our notation and introduce
basic concepts to be used in the rest of the paper.
Questions regarding ensemble definitions, for CDRG-m as well as
CDRG-s, will be discussed in section \ref{Models}.
Section \ref{Config} contains a basic statistical analysis of the
ensembles as seen from the point of view of the stubs.
In section \ref{Local}, we will discuss the statistics of the number
of copies of an arbitrary small graph as a subgraph of a random graph,
and define rules for the computation of the asymptotically expected
counts, pointing out differences and similarities between CDRG-s and
CDRG-m ensembles.
In section \ref{Global}, we will discuss the global properties of
random graphs from CDRG ensembles, as revealed by a generating
function analysis of the cluster size distribution, extending the
analysis presented in ref. \cite{CDRG}.
Both the global and local analysis reveal a certain redundancy
(symmetry) property of CDRG models, which forms the subject of section
\ref{Redundancy}.
In section \ref{Relations} we will identify subclasses of CDRG
ensembles corresponding to commonly studied models.
Section \ref{Conclusion}, finally, contains a resum\'e of our main
results, and some concluding remarks and speculations.

\section{Notation and basic concepts}
\label{Notation}

A {\em labelled graph} consists of a set of distinguishable {\em
vertices} (nodes, sites, points), which may be pairwise connected by
{\em edges} (links, bonds, lines).

Unless otherwise stated, a graph is assumed to be {\em symmetric}
(undirected), such that edges have no particular direction (as opposed
to a {\em digraph} -- or directed graph -- where an edge has a
direction, pointing from one vertex to another).

A graph with $N$ vertices is conveniently represented by its symmetric
$N\times N$ {\em adjacency matrix} $\S$. An element $S_{ij}$ counts
the number of edges between vertices $i$ and $j$; thus, each edge
contributes both to $S_{ij}$ and $S_{ji}$; as a result each diagonal
element $S_{ii}$ will be even, representing twice the number of
self-couplings of vertex $i$.  In a {\em simple} graph, cycles of
length one (self-couplings or tadpoles) and two (multiple edges) are
absent; as a result, the diagonal elements of $\S$ are zero, and the
remaining elements are restricted to the values zero or one. A {\em
multigraph} may be simple or degenerate.

The {\em degree} (or connectivity) $m$ of a vertex is defined as the
number of edges connected to it, given by the corresponding row sum
$\sum_j S_{ij}$; the vertex can be considered as possessing $m$ {\em
stubs} -- points where a single edge endpoint ({\em butt}) is
attached.

It is sometimes convenient to consider not only the vertices, but also
the edges, and indeed the individual stubs and butts, as being
distinguishable.


The {\em degree sequence} of a graph is an ordered list of $N$
integers $(m_1\dots m_N)$, describing the individual degrees of the
$N$ vertices.  Alternatively, it can be summarized in terms of the
{\em degree counts}, $N_m=\sum_i \delta(m,m_i)$, counting the number
of vertices having degree $m$.

A commonly studied class of ensembles is based on giving an asymptotic
{\em degree distribution} $\{p_m\}$, from which a compatible degree
sequence can be determined for a given graph size $N$ with degree
counts $N_m\approx N p_m$. Then a random compatible graph is chosen by
means of a random stub pairing (the {\em configuration model}
\cite{BenCan,MoRe98}).  This approach will be referred to as {\em
DRG}, for degree-based random graphs.

In another approach, {\em IRG} (for inhomogeneous random graphs), a
class of vertex-colored extensions of the classic model has been
considered, where each vertex is randomly and independently assigned
an abstract type ({\em color}) drawn from a given distribution, and
where edge probabilities are allowed to depend on the connected pair
of colors \cite{Sod02}.

In a recent article \cite{CDRG}, the philosophies behind DRG and IRG
were combined in a novel approach, where a hidden {\em stub-coloring}
was used to define a very general class of ensembles with a given
degree distribution.
This approach, CDRG, forms the main subject of this article.

Thus, we will consider {\em stub-colored} graphs, where each stub
independently carries an internal characteristic, a {\em hidden color}
$a\in [1,\dots,K]$, to be considered unobservable.  The degree $m$ of
a vertex then decomposes into the sum of contributions $m_a$ counting
the stubs with a definite color $a$. These sub-degrees can be
collected in a $K$-vector $\m=(m_1\dots m_K)$, to be referred to as
the {\em colored degree} of the vertex.

It is then natural to consider the {\em colored degree sequence} of
such a graph, in terms of the numbers $N_{\m}$ of vertices with a
distinct colored degree $\m$.

Accordingly, each edge connects a pair of colored stubs and can be
associated with a color pair $(a,b)$.
We can then also consider the count $n_{ab} = n_{ba}$ of edges for
each color pair, where an $ab$-edge for practical reasons contributes
both to $n_{ab}$ and $n_{ba}$ (so diagonal elements $n_{aa}$ are
even).

The total number of butts with color $a$ in the graph is then given by
$\sum_b n_{ab}$; this must match the corresponding stub count $M_a
\equiv\sum_{\m} m_a N_{\m}$.
In particular, the total butt count, $\sum_{ab} n_{ab}$, must be even
(being twice the number of edges), and it must equal the total stub
count, $M = \sum_a M_a=\sum_{\m} \sum_a m_a N_{\m}$.
We will find it convenient to collect the colored stub counts in a
vector $\M = (M_1,\dots,M_K)$.

Throughout this article, $K$-vectors will be denoted by (mostly lower
case, with $\M$ being an exception) fat symbols such as $\x \equiv
(x_1\dots x_K)$, in terms of which an obvious simplified notation will
be used: $\x^{\m}\equiv \prod_a x_a^{m_a}$, $\m!\equiv\prod_a m_a!$,
etc. The uniform $K$-vector $(1,\dots 1)$ will be denoted as
$\1$. Similarly, $K\times K$-matrices will be denoted by upper case
fat symbols such as $\T=\{T_{ab}\}$, with matrix product indicated by
juxtaposition. A {\em componentwise} product will be denoted by a
cross ($\times$), as in $\x\times\m\equiv\(x_1 m_1,\dots,x_K
m_K\)$. The transpose of a matrix $\T$ will be denoted by $\t{\T}$ and
the matrix inverse of the transpose by $\t{-\T}$.

We will be interested in models based on a definite {\em colored
degree distribution} (CDD) $\{p_{\m}\}$, in terms of which we can
define moments $\ave{m_a}=\sum_{\m}p_{\m}m_a$, etc.  Such a distribution
is conveniently described by its multivariate {\em generating
function},
\beq
\label{H}
	H(\x)=\sum_{\m}p_{\m}\x^{\m},
\eeq
where $\x=(x_1\dots x_K)$ is a $K$-component vector of auxiliary
variables.

From $H$ the invidual $p_{\m}$ can be extracted by means of repeated
differentiation at $\x=\0$, while repeated differentiation at $\x=\1$
yields the {\em combinatorial moments}
\beq
\label{E}
	E_{ab\dots}=\partial_a \partial_b\dots H(\x=\1),
\eeq
where $\partial_a$ stands for the derivative with respect to $x_a$.
Thus, the lowest moments become $E_a=\ave{m_a}$,
$E_{ab}=\ave{m_am_b-m_a\delta_{ab}}$, etc., generalizing the
corresponding combinatorial moments of the {\em total} degree,
$\ave{m}$, $\ave{m(m-1)}$, etc. Occasionally we will suppress indices
and refer to the $n$th order combinatorial moment as $\E_{(n)}$. Thus,
$\E_{(1)}=\{E_a\}=\ave{\m}$, $\E_{(2)}=\{E_{ab}\}$,
$\E_{(3)}=\{E_{abc}\}$, etc.
In particular, it is frequently convenient to view the second order
tensor $\E_{(2)}$ as a {\em matrix}, denoted simply by $\E$.

Upon summing over the indices independently, the $n$th order {\em
scalar} combinatorial moments result, denoted $E_{(n)}$. Thus,
$E_{(1)}=\sum_a E_a = \sum_a \ave{m_a} = \ave{m}$, $E_{(2)}=\sum_{ab}
E_{ab} = \ave{m(m-1)}$, $E_{(3)}=\sum_{abc}
E_{abc} = \ave{m(m-1)(m-2)}$, etc.

\section{Model definitions}
\label{Models}

Ensembles in CDRG are based on {\em asymptotic models}, where a
desired asymptotic behaviour as $N\to\infty$ is specified.  For a
given asymptotic model, finite graph ensembles can be defined.

\subsection{Asymptotic CDRG model}

An asymptotic model is defined as follows.
\\ \\ \fbox{\parbox{76mm}{ {\bf Asymptotic CDRG model:}\\
\begin{itemize}
\item Specify the desired color space, taken to be $[1,\dots,K]$ for
some integer $K\ge 1$;

\item Choose a normalized asymptotic colored degree distribution
$\{p_{\m}\}$, with $p_{\m}\ge 0$ and $\sum_{\m} p_{\m} = 1$;
\item Choose a symmetric $K\times K$ {\em color preference matrix}
$\T$, with real, non-negative elements $T_{ab}\ge 0$, subject to the
constraint
\beq
\label{Tconstr}
	\sum_b T_{ab} \ave{m_b} = 1.
\eeq
\end{itemize}
}}
\\ \\
The role of $\T$ is to control the asymptotic symmetrized
color-specific distribution of edges: $n_{ab} \sim
N\ave{m_a}T_{ab}\ave{m_b}$, where $n_{ab}$ denotes the number of edges
connecting colors $a$ and $b$. The constraint (\ref{Tconstr}) is
needed for the mutual consistency between the asymptotic vertex and
edge statistics -- roughly speaking, it secures a matching butt for
each stub.

Following ref. \cite{CDRG}, we will for simplicity assume colored
degree distributions to be well-behaved, such that all moments of
arbitrary order are defined.
This excludes power tails in the degree distribution -- the particular
complications associated with extending CDRG to {\em fat-tailed}
distributions fall outside the scope of this article, and will
hopefully be the subject of a future paper.

\subsection{Ensembles of finite graphs}

Based on a given asymptotic model, we wish to define an ensemble of
multigraphs or simple graphs with a given size $N$.

\subsubsection{Multigraph ensembles -- CDRG-m}

The simplest and most straightforward way to define an ensemble of
{\em multigraphs} of a given size $N$ consistently with a given
asymptotic CDRG model is as follows. Fix the color-specific vertex and
edge counts, $N_{\m}$ and $n_{ab}$, as close as possible to their
expected values, i.e. $N_{\m} \approx N p_{\m}$, and $n_{ab} \approx N
\ave{m_a} T_{ab} \ave{m_b}$, such that they yield matching colored
stub and butt counts, $\sum_{\m} N_{\m} m_a = \sum_b n_{ab} = M_a
\approx N\ave{m_a}$. Then place edges by for each color $a$ randomly
pairing each of the $M_a$ stubs with a unique matching butt.

The result can be considered a {\em microcanonical ensemble} of
multigraphs, and was used in the original article \cite{CDRG} as a
means to define an ensemble of simple graphs by projecting out the
simple part.

In this article, we will consider
a slightly different multigraph ensemble where only $N$ is fixed while
the other counts are allowed to vary.
While being slightly more elaborate to implement as a random graph
generator, this {\em grand canonical} ensemble is more convenient for
analytical purposes.
\\ \\ \fbox{\parbox{76mm}{ {\bf Grand canonical multigraph ensemble}\\
\begin{enumerate}
\item For each of the $N$ vertices, draw its colored degree at random
from the asymptotic distribution $\{p_{\m}\}$. The result is a random
colored degree sequence, yielding a definite stub count $M$, the
expected value of which is $N\ave{m}$. Repeat this step until $M$ is
even.
\item Consider the entire set of $(M-1)!!$ pairings of the $M$ stubs,
and associate with each pairing a statistical weight given by the
product of single edge factors, where each $ab$-edge contributes a
factor $T_{ab}/N$.  Draw a pairing at random from the resulting
weighted distribution.
\end{enumerate}
}} \\ \\
The weighted random pairing defines a natural colored extension of the
stub-pairing method, the configuration model, as used in DRG
\cite{MoRe98}.

In the thermodynamic limit, the microcanonical and grand canonical
ensembles corresponding to the same asymptotic model should be
statistically equivalent. Indeed, when $N\to\infty$, the distribution
of colored degree counts $N_{\m}$ in the grand canonical ensemble
becomes sharply peaked around the microcanonical values $\ave{N_{\m}}
= Np_{\m}$.
A a result, the total colored stub counts $M_a$ will be close to
$N\ave{m_a}$, and as will be shown below, this implies that the
distribution of colored edge counts $n_{ab}$ resulting from the
weighted pairing becomes sharply peaked around the microcanonical
ensemble values $\ave{n_{ab}}=N\ave{m_a}T_{ab}\ave{m_b}$.
In the next section we will give a detailed analysis of the basic stub
pairing statistics.

\subsubsection{Simple graph ensembles -- CDRG-s}

In ref. \cite{CDRG}, a microcanonical ensemble of simple graphs was
defined by projecting out the simple graph part from the
microcanonical ensemble of multigraphs, as realized by redoing the
random butt-stub pairing step until a simple graphs results.

Here, we shall instead consider a {\em grand canonical ensemble} of
simple graphs, defined by projecting out the simple part from the
corresponding CDRG-m ensemble. It can be realized e.g. by repeatedly
drawing a member of the latter until a nondegenerate graph results.

The efficiency of this method depends on the probability for a
randomly drawn multigraph to be simple. This probability is easily
computed, as will be demonstrated below (in the section on local
characteristics), where we will verify the result given in
ref. \cite{CDRG}.

In ref. \cite{CDRG} it was also argued that several statistical graph
properties not directly involving the presence or absence of
degeneracies as measured in a CDRG ensemble of simple graphs were
asymptotically identical to those of the underlying multigraph
ensemble; we shall provide arguments that this is indeed the case.

\section{Basic stub statistics}
\label{Config}

For the forthcoming analysis of local and global structural properties
of random graphs drawn from the grand canonical ensemble of
multigraphs, an initial basic statistical analysis of the graph
properties as seen from the point of view of the individual stubs is
required.

\subsection{Colored stub distribution}

In a grand canonical CDRG-m enesemble, each vertex $i$ can be
considered to have an independent random colored degree $\m_i$ drawn
from the asymptotic distribution $\{p_{\m}\}$ (neglecting the slight
modification due to the constraint of even $M$). Hence, the vector
$\M=\sum_i \m_i$ of total colored stub counts is essentially the sum
of $N$ independent colored degrees, which trivially results in the
$\M$ distribution $P_{\M}$ being centered around the expected stub
count $\ave{\M} = N\ave{\m}$, with fluctuations of $O(N^{1/2})$
governed by the correlation matrix $\ave{\M \t{\M}}_c = N \ave{\m
\t{\m}}_c$.

For the derivation of more general properties of $P_{\M}$, it may be
convenient to use its generating function, which is given by $H(\z)^N
= \sum_{\M} P_{\M}\z^{\M}$,
\footnote{Due to the restriction to even $M$, the generating function
is really given by $\(H(\z)^N + H(-\z)^N\)/\(H(\1)^N + H(-\1)^N\)$,
yielding an unimportant modification. We will use $H(\z)^N$ for
simplicity, keeping in mind that only the even-$M$ part is physical.}
where $H(\z)$ is the generating function for $p_{\m}$, as defined in
\eq{H}. From $H(\z)^N$, $P_{\M}$ can be extracted as the coefficient
for $\z^{\M}$:
\beq
\label{PM}
	P_{\M} = \oint\D{\z} \z^{-\M} H(\z)^N,
\eeq
where $\oint\D{\z}$ stands for $\prod_a\oint\D{z_a}$, denoting the
complex integration of each $z_a$ along a path encircling the origin.
For $\M$ close to its average $N\ave{\m}$, the integral is
asymptotically dominated by the contributions from a saddlepoint
$\z\approx\1$, from which the asymptotic properties of $P_{\M}$ can be
derived in a saddlepoint approximation.

\subsection{Stub pairing statistics}

Next, we wish to analyze the result from the weighted random pairing
of stubs.  To that end we note that for a given assignment of colored
vertex degrees, the only thing important for the pairing step is the
total stub count $\M=\{M_a\}=\sum_i \m_i$.

Denote by $Z(\M)$ the total weight of the set of $(M-1)!!$ possible
stub pairings, given $\M$.
It is the sum over distinct pairings $\pi$ of the associated product
of edge weights $T_{ab}/N$, and can be written as follows:
\begin{subequations}
\label{ZM}
\beqa
\nonumber
	&&Z(\M)=\sum_{\pi} \prod_{\text{pairs}} \frac{T_{ab}}{N}
\\
\label{Zn}
	&=&
	N^{-M/2} \M! \sum_{\{n_{ab}\}}
	\prod_{a<b} \frac{T_{ab}^{n_{ab}}}{n_{ab}!}
	\prod_a \frac{T_{aa}^{n_{aa}/2}}{n_{aa}!!}
\\
\label{Zz}
	&=&
	N^{-M} \M! \oint \D{\z} \z^{-\M} e^{\frac{N}{2} \t{\z} \T \z},
\eeqa
\end{subequations}
where the sum over $\{n_{ab}\}$ is restricted to non-negative,
symmetric values with even diagonal and correct row sums, $\sum_b
n_{ab} = M_a$. The last form, (\ref{Zz}), is obtained by
Fourier-expanding the implicit Kronecker deltas for the row sum
constraints.

So far, everything is exact. The complex integral form of $Z(\M)$ can
be estimated in a saddlepoint approximation, based on extremizing the
associated ``action'', $S(\z) = \M\cdot\log(\z) - \frac{N}{2} \t{\z}
\T \z$. Demanding a vanishing derivative, $\partial_{z_a} S =
\frac{\M}{\z} - N\T\z = 0$, yields the equation for a saddlepoint as
\beq
	\M = N \z \times\(\T \z\),
\eeq
implicitly defining the saddlepoint $\z(\M)$ (up to a total sign,
really, but for even $M$, the two yield identical contributions).

For the particular choice of $\M = N\ave{\m}$, defining the expected
value of $\M$, the relevant solution is $\z=\ave{\m}$, yielding for
the total weight the asymptotic value $Z\(\M=N\ave{\m}\) \sim
e^{-N\ave{m}/2}$, where we have disregarded subexponential factors and
assumed $M$ to be even.
The value of $Z(\M)$ for slightly different arguments can then be
estimated by noting that a small relative change in $\M$ yields a
small relative change in $\z$, and leads to a small change in the
value of the action $S$.

Thus, upon replacing $\M$ by a modified value $\hat{\M}=\M+\epsilon$,
the saddlepoint $\z$ changes to $\hat{\z}=\z+\delta$, and the action
$S=S(\M,\z)$ changes to $\hat{S}=S + \epsilon \cdot \partial
S/\partial \M + \delta \cdot \partial S/\partial \z$, evaluated at
$\M=N\ave{\m}$, $\z=\ave{\m}$, where the $\z$ derivative vanishes due
to the saddlepoint condition. Thus, to lowest order, the modified
value of the action is given by $\hat{S}=S+\epsilon\cdot\log(\z)$. As
a result, the complex integral to leading order changes by a factor of
$\z^{-\epsilon}$, and thus the total weight $Z$ changes by a factor
$(\M/N\z)^{\epsilon} \approx 1$ -- i.e. not at all.  This means that
$Z(\M)$ has a saddlepoint for $\M$ close to its expected value,
$\ave{\M}=N\ave{\m}$.

\subsection{Individual pairing probabilities}

The asymptotic probability that an arbitrarily chosen pair of stubs
will be connected in the random pairing, given their colors $a,b$, can
be calculated as the ratio of the total weight {\em conditional} on
this connection and the unconditional total weight. The conditional
weight is obtained by multiplying the factor $T_{ab}/N$ for the
clamped edge by the total weight $Z\(\M - \e_a - \e_b\)$ of all
pairings of the remaining $M-2$ stubs, where $\e_a$ denotes the unit
vector along the positive $a$-direction. This is to be divided by
$Z(\M)$; as argued above the $Z$ ratio is asymptotically $1$, and so
the asymptotic probability is simply $T_{ab}/N$.

Let us check this result for consistency: There are $M_b$ stubs with
color $b$; each of these defines an equally probable matching partner
to a fixed stub of color $a$ (neglecting for the case $a=b$ the
asymptotically negligible possibility that the two stubs be
identical), yielding $T_{ab} M_b/N \approx T_{ab}\ave{m_b}$ for the
probability that the pairing partner of an arbitrary stub of color $a$
has color $b$. A final summation over $b$ yields $\sum_b
T_{ab}\ave{m_b} = 1$, expressing the correct normalization of the
asymptotic probabilities.

The argument is easily extended to yield the asymptotic probability
for an arbitrary finite number of clamped stub pairs in the grand
canonical ensemble of multigraphs, as given simply by the product of
the corresponding edge factors $T_{ab}/N$, with the relative error
being of order $O\(N^{-1}\)$.

From these pairing probabilities we can draw the trivial conclusion
that the colored edge counts $n_{ab}$ in a grand canonical CDRG-m
ensemble asymptotically will be close to the corresponding
microcanonical ensemble values $N \ave{m_a} T_{ab} \ave{m_b}$; this
can also be derived directly from eqs. (\ref{ZM}).

Conversely, it is easily realized that asymptotically identical
pairing probabilities hold for the microcanonical multigraph ensemble,
where the colored edge counts are fixed to $n_{ab}\sim
N\ave{m_a}T_{ab}\ave{m_b}$. Given an arbitrary pair of distinct stubs
with respective colors $a,b$, the probability that they be paired is
the product of (1) the probability $n_{ab}/M_a$ that the first stub is
chosen to belong to the group of $a$-stubs selected to be paired with
color $b$, (2) the corresponding probability $n_{ab}/M_b$ for the
other stub, and (3), the probability $1/n_{ab}$ that the first stub is
paired with the second among the $n_{ab}$ candidates. Multiplying the
three factors together yields the probability $n_{ab}/(M_aM_b) \sim
T_{ab}/N$.

\section{Local characteristics}
\label{Local}

Calculability of local as well as global graph characteristics in a
model greatly simplifies the task of model inference from observed
graphs.  All local graph characteristics can be derived from the
embedding counts of various small connected {\em subgraphs}. These are
easy to measure in observed graphs. The analysis given in the previous
section provides the necessary tools for deriving rules for
calculating the asymptotically expected count distributions in a CDRG
model.  We will first consider the case of a CDRG-m; the results for
that case can then be used to derive the corresponding results for
CDRG-s. Except where otherwise stated, the grand canonical ensembles
will be assumed.

\subsection{Subgraph statistics I: Multigraph ensemble}

\subsubsection{Initial discussion}

Given an arbitrary, possibly degenerate, small graph $\g$ with $v$
vertices and $e$ edges, we wish to study the statistics of the number
$n_{\g}$ of distinct copies of $\g$ found in a random graph $\G$ drawn
from a CDRG-m ensemble, i.e. the number of distinct subgraphs of $\G$
isomorphic to $\g$.

A subgraph of $\G$ is defined as a subset $\vv$ of the $N$ vertices of
$\G$, together with a subset $\ee$ of the edges among $\vv$.  Two
subgraphs are considered distinct if they have different $\vv$ or
different $\ee$.
Note that a general subgraph is not necessarily an {\em induced}
subgraph, where $\ee$ must be the entire set of edges among $\vv$.
Thus, e.g., if $\g$ lacks an edge between a pair of vertices, the
corresponding pair in the target set $\vv$ may well be connected.

We are primarily interested in connected $\g$, but we will allow
ourselves to consider also cases where $\g$ is not connected.  Let us
begin by considering a few simple examples explicitly.

{\bf Single vertex ($\bullet$):} Let $\g$ be a {\em single vertex with
no edges}. Then we must have $n_{\g}=N$, since there are $N$ ways to
choose a single target vertex in $\G$.

{\bf Unconnected pair of vertices ($\bullet\;\bullet$):} Let $\g$
consist of {\em two vertices and no edges} (so $\g$ is not
connected!). Then, $n_{\g}=N(N-1)/2\sim N^2/2$, reflecting the
$N(N-1)$ ways to choose an ordered pair of vertices in $\G$, while the
symmetry of $\g$ under interchange of the two vertices makes the two
{\em a priori} distinct orderings equivalent.

{\bf Connected pair ($\bullet$--$\bullet$):} Let $\g$ be the graph
consisting of two vertices connected by a single edge. Again, $\g$ is
symmetric under interchange of its two vertices, and the target pair
$\vv$ of vertices can be chosen in $N(N-1)/2$ distinct ways. Not all
vertex pairs are connected, while others are multiply connected: A
pair with $k$ connections yields $k$ distinct copies of $\g$.  The
average number of connections between an arbitrary pair of vertices is
the sum over color pairs $a,b$ of the average number of $ab$-edges
connecting them. Each vertex of the pair has a colored degree randomly
drawn from $\{p_{\m}\}$, For a given pair $\m,\m'$, there are
$m_am'_b$ possible ways to choose the $a,b$-edge, each yielding a
probability $T_{ab}/N$.  Averaging this over the colored degrees
$\m,\m'$ yields $\sum_{\m} p_{\m}\sum_{\m'} p_{\m'} m_a m'_b T_{ab}/N
= \ave{m_a} \ave{m_b} T_{ab}/N$.  Finally, summing over $a,b$ gives
$\t{\ave{\m}} \T \ave{\m} / N = \ave{m} / N$ for the expected number
of edges between a randomly chosen pair of vertices. Multiplying this
by the number of ways to choose the pair of vertices yields
$\frac{N}{2}\ave{m}$ for the asymptotically expected number of copies;
this is precisely the expected number ef edges, $\ave{M}/2$, as it
must be, since every edge defines a distinct copy of $\g$.

\subsubsection{Expected count for general $\g$}
\label{Feynman}

For a more general graph $\g$, the expected count can be computed by
multiplying the number of ways $\binom{N}{v}$ to choose the vertex
target set $\vv$ by the expected number of copies using a fixed target
set $\vv$. The latter is obviously independent of $\vv$ when $\G$ is a
random graph, and is the sum of the expected number of copies for each
of the (naively $v!$) inequivalent orderings of $\vv$, defined as the
number of ways to choose the target set $\ee$ from the existing edges
among $\vv$ (e.g., an ordered $k$-tuple of edges between a specific
pair of vertices in $\vv$ can be chosen in $n!/(n-k)!$ distinct ways,
if the target pair in $\vv$ is connected by $n$ edges).

In addition, if $\g$ has a nontrivial isomorphism group (in terms of
permutations of vertices as well as permutations and flips of edges),
the result must be divided by a {\em symmetry factor} $S_{\g}$, given
by the order of this group. It consists in two factors: One is given
by the order of the {\em vertex} permutation symmetry of $\g$, the
other by the order of the group of permutations and flips of {\em
edges} with fixed vertices leaving $\g$ invariant, yielding a factor
of $n!$ for each pair of distinct vertices in $\g$ connected by $n$
edges, and a factor of $n!2^n=(2n)!!$ for each vertex with $n$
tadpoles (requiring $2n$ stubs).

This results in the following {\em ``Feynman'' rules} for the
calculation of the asymptotically expected number $n_{\g}$ of copies
of an arbitrary small graph $\g$ in a large random graph $\G$ drawn
from a CDRG-m ensemble:
\\ \\ \fbox{\parbox{76mm}{ {\bf Rules for calculating expected
asymptotic subgraph counts $\ave{n_{\g}}$:}\\
\begin{enumerate}
\item Label each stub in $\g$ with an independent color index;
\item Associate with every vertex in $\g$ with $n$ stubs labelled
$a,b,\dots$ a factor given by $N$ times the corresponding component
$E_{ab\dots}$ of the $n$th order combinatorial moment $\E_{(n)}$;
\item Associate with each edge in $\g$ a factor $T_{ab}/N$, where
$a,b$ are the color labels of the connected stubs;
\item Multiply together all vertex and edge factors, and sum the
result over the stub colors.
\item Divide the result by the proper symmetry factor $S_{\g}$, to
yield the expected count $\ave{n_{\g}}$.
\end{enumerate}
}} \\ \\
{\bf Sketch of proof}: The individual {\em vertex factor} decomposes
into a factor of $N$ for the number of ways to choose the target
vertex in $\G$, and a factor $E_{ab\dots}$, which takes some
explaining. Consider a vertex in $\g$ with two stubs, assigned colors
$a,b$. The colored degree $\m$ of the target vertex is drawn from
$p_{\m}$, and the number of ways to pick two stubs with correct
colors, given $\m$, is $m_a m_b$ if $a\neq b$, and $m_a(m_a-1)$ if
$a=b$. Averaging over $\m$ yields $E_{ab}$. The result generalizes to
an arbitrary number of stubs.

The {\em edge factor} $T_{ab}/N$ represents the individual stub-stub
connection probability as derived in the previous section; it
ultimately stems from the corresponding factor in the weighted random
pairing involved in the definition of the grand canonical ensemble.

The vertex part of the {\em symmetry factor} simply stems from the
fact that the existence of a vertex permutation symmetry of $\g$
implies a reduction of the naive number $N(N-1)\dots(N-v+1)\sim N^v$
of inequivalent choices of ordered target sets $\vv$. Similarly, the
edge part reflects the equivalence of naively distinct edge target
sets $\ee$ for the same $\vv$, differing only by the interchange of
edges connecting the same pair of vertices, or by a flip of a single
edge connecting a vertex to itself.

The same asymptotic rules can be derived for the case of the
microcanonical multigraph ensemble using similar arguments.

In table \ref{sub}, the expected counts are given for subgraphs in the
form of {\em chains}, {\em stars} and simple {\em cycles} of arbitrary
length for a CDRG-m model, and for a plain DRG model (CDRG-m
restricted to a single color) for comparison.
\begin{table*}{!}
\centering
\begin{tabular}{|c|c|c|c|c|c|c|c|}
\hline
Subgraph      & $k$   & Vertices & Edges & Diff. & Symm. factor &  $\ave{n_{\g}}_{\text{CDRG}}$ & $\ave{n_{\g}}_{\text{DRG}}$  \\
         $\g$ &  range   & $v$   & $e$   & $v-e$ & $S_{\g}$     &                               &                              \\
\hline
$k$-star\rule[-3mm]{0mm}{8mm} & $k\ge 2$ & $k+1$ & $k$   & 1     & $k!$     & $N E_{(k)} / k!$                 & $N E_{(k)} / k!$                  \\
\hline
$k$-chain\rule[-3mm]{0mm}{8mm}  & $k\ge 2$ & $k$   & $k-1$ & 1     & $2$      & $\frac{N}{2}\t{\1}\E(\T\E)^{k-3}\1$ & $\frac{N}{2} E\(E/\ave{m}\)^{k-3}$ \\
\hline
$k$-cycle\rule[-3mm]{0mm}{8mm} & $k\ge 3$ & $k$   & $k$   & 0     & $2k$     & $\tr\(\T\E\)^k / (2k)$           & $\(E/\ave{m}\)^k/(2k)$           \\
\hline
\end{tabular}
\caption{Asymptotically expected counts of subgraphs in the form of
{\em stars}, {\em chains}, and simple {\em cycles} of arbitrary size
as computed in a CDRG-m model and, for comparison, in a corresponding
uncolored (DRG) model.
The {\bf $k$-star} consists of a single ``hub'' vertex connected to
each of $k$ leaf nodes by a single edge. The symmetry factor of $k!$
is due to permutations of the $k$ leaves. The factors for the $k$
leaves have been simplified as described in the text.  For both CDRG
and DRG, the resulting expected count can be written as $N
\ave{\binom{m} {k}}$, as may have been expected - each vertex in $\G$
with $m\ge k$ stubs defines $\binom{m}{k}$ copies; in this case the
expected count depends only on the plain degree distribution,
$\{p_m\}$.
The {\bf $k$-chain} consists of $k$ vertices connected into a chain
by $k-1$ edges. The symmetry factor of $2$ is due to a flip of the
entire chain. The two leaf factors for the endpoints have been
simplified.  As a result, the 4-chain is the first chain where the
expected count shows a nontrivial dependence on $\T$, distinguishing
CDRG from plain DRG for which the expected chain counts form a simple
geometric series.
The {\bf $k$-cycle} consists of $k$ vertices connected into a
closed loop by $k$ edges. The symmetry factor $2k$ is due to flipping
($\to 2$) and rotating ($\to k$) the vertex order in the cycle.  }
\label{sub}
\end{table*}
Note the simplification occurring in the expression for the expected
count for each {\em leaf} node with a single connection, due to the
identity $\T\ave{\m}=\1$: The vertex factor for the leaf and the single
edge factor gives upon summation of the color label assigned to the
single stub a factor $\sum_a N \ave{m_a}\times T_{ab}/N \equiv 1$, and
their only effect is to increase the degree of the moment associated
with the neighboring vertex by adding an index ($b$) that is simply
summed over.

\subsubsection{Scaling with $N$ and edge correlations}

Of obvious interest is how $n_{\g}$ scales with $N$. The rules for the
calculation of the expected count yield a factor of $N$ for each
vertex and a factor of $N^{-1}$ for every edge, so the total power of
$N$ is $v-e$, which can also be expressed as {\em the number of
mutually disconnected components in $\g$, minus the number of loops in
$\g$}. For a connected $\g$, this yields 1 minus its number of
loops. Thus if $\g$ is a {\em tree}, the expected number of copies
scales as $O(N)$, while for a {\em one-loop connected} $\g$ the
expected number scales as $O(1)$; for any connected $\g$ with more
than one loop there are asymptotically no copies at all, since the
expected number is suppressed by factors of $N$.

Let us demonstrate with a few simple examples the increased
correlation possibilities in CDRG models as opposed to a plain DRG
model. First, we compare the counts for {\em triangles} (3-cycles,
$\Delta$), {\em wedges} (3-chains, $\Lambda$), and {\em edges}
(2-chains, $I$). In a CDRG ensemble, their respective expected counts
are
\begin{subequations}
\beqa
	\ave{n_{\Delta}} &=& \frac{\tr(\T\E)^3}{6},
\\
	\ave{n_{\Lambda}} &=& \frac{NE}{2},
\\
	\ave{n_{I}} &=& \frac{N\ave{m}}{2}.
\eeqa
\end{subequations}
A {\em plain DRG} ensemble yields as identical expression for
$\ave{n_{\Lambda}}$ as well as for $\ave{n_{I}}$, while the triangle
count becomes $\ave{n_{\Delta}}_{\text{DRG}}=E^3/(6\ave{m}^3)$,
yielding the relation
\beq
	\ave{n_{\Delta}}_{\text{DRG}} = \frac{\ave{n_{\Lambda}}^3}{6\ave{n_{I}}^3},
\eeq
absent in a generic CDRG ensemble.

Similarly, the expected {\em k-chain} count is
$\ave{n_k}=N\t{\1}\E\(\T\E\)^{k-3}\1/2$. In plain DRG, this simplifies
to a geometric series, $NE^{k-2}/(2\ave{m}^{k-3})$, which again can be
expressed in terms of the wedge and edge counts:
\beq
	\ave{n_k}_{\text{DRG}} = \ave{n_{\Lambda}}^{k-2} \ave{n_{I}}^{3-k},
\eeq
whereas in CDRG, this strict relation is absent.

A popular edge correlation measure in the literature is the so called
{\em clustering coefficient} $C$, defined as the probability that two
randomly chosen neighbors of a random vertex are connected
\cite{WatStr98,MasSne02}. In not-so-sparse random graph models with an
excessive amount of triangles, as can be anticipated to result with a
power tail in the degree distribution, or in models based on an
underlying regular structure, $C$ can attain a finite value.  This is
not the case in the type of models we are considering here, and we
expect $C$ to decrease as $O(N^{-1})$.  We can estimate $C$ by
comparing the expected counts for triangles (3-cycles) and
3-chains. Their ratio multiplied by 3 gives the estimate $C =
\tr(\T\E)^3/(N E)$. While this indeed scales as $O(N^{-1})$, the
finite number $NC$ has a nontrivial dependence on $\T$, allowing it to
deviate from the DRG value of $E^2/\ave{m}^3$.

These examples serve to illustrate the role of the hidden color in
enabling a non-trivial edge correlation structure, and in lifting the
simple relations between different subgraph counts present in DRG.

\subsubsection{Beyound expected counts: Distribution shape}

The expected count $\ave{n_{\g}}$ of a given subgraph $\g$
gives only partial information on the count distribution.
Of interest are also the actual shapes of the count distributions, as
well as the correlations between different subgraph counts.

A first step in this direction is given by considering the expected
{\em squared count}, $\ave{n_{\g}^2}$, for a fixed graph $\g$.  The
count itself consists in a sum over embedding positions, and so the
squared count is given by summing over two independent embedding
positions, which can be reorganized as a sum over the {\em relative}
position of the two copies as defined by vertex and edge coincidences,
and a sum over the {\em absolute} embedding position of the resulting
composite graph.

The key point is that the contribution to $\ave{n_{\g}^2}$ from
each possible configuration of the composite graph $\g_2$ is given by
{\em its naive expected count $\ave{n_{\g_2}}$ as a subgraph of $\G$,
multiplied by the number of distinct ways to combine the two copies
into $\g_2$}.
The multiplication by the number of ways to obtain $\g_2$ compensates
e.g. for the extra twofold symmetry typically arising in $\g_2$,
related to the interchange of the two copies.

Let us consider the case of a connected $\g$ being a tree or having a
single loop, and do a brief analysis of the possible scaling
properties of the expected count $\ave{n_{\g_2}}$ of the combined
graph $\g_2$.

For a connected $\g$, the expected embedding count scales as
$O(N^{v-e})$, yielding $O(N)$ for a tree and $O(1)$ for a one-loop
graph. When combining two copies of $\g$ into $\g_2$, they may overlap
in a common subgraph, the {\em overlap graph}, with edge and vertex
counts $e_o,v_o$.  Then the combined graph $\g_2$ will have vertex
count $v_2=2v-v_o$ and edge count $e_2=2e-e_o$, and its expected count
will scale as $O(N^{2v-2e-v_o+e_o})$.

If $\g$ is a {\em tree}, its only possible overlap graphs are forests,
with $v_o-e_o\ge 0$, with equality only for the empty subgraph.  This
means that the leading $O(N^2)$ contribution to $\ave{n_{\g}^2}$ comes
entirely from the case where the two copies of $\g$ are completely
disjoint, yielding a leading contribution to $\ave{n_{\g}^2}$ matching
that of $\ave{n_{\g}}^2$, while the remaining contributions scale at
most as $O(N)$.  As a result the standard deviation of the $\g$ count
scales at most as $O(N^{1/2})$, as compared to the $O(N)$ behavior of
the expected count, yielding an asymptotically sharp distribution for
the corresponding {\em intensive} entity, the {\em count density}
$\rho_{\g}=n_{\g}/N$.

If $\g$ has a {\em single loop}, we have $v-e=0$, and the expected
count is {\em finite}. Then we are interested in contributions to
$\ave{n_{\g}^2}$ scaling at least as $O(1)$, requiring $v_o-e_o\le 0$.
Hence, the only interesting overlap graphs between the two copies of
$\g$ are the {\em empty graph} and connected one-loop graphs
(including the entire $\g$) where the two copies of $\g$ share the
loop part (possibly rotated or flipped) both yielding $v_o-e_o=0$.
There are two possibilities here.

If $\g$ consists of a {\em bare loop} without decorations, the only
interesting contributions to $\ave{n_{\g}^2}$ are those from cases
where the two copies are completely disjoint or completely identical,
yielding $\ave{n_{\g}^2}=\ave{n_{\g}}^2+\ave{n_{\g}}$ to leading
order. The argument can be generalized to higher moments of $n_{\g}$,
showing that the asymptotic distribution of the $n_{\g}$ is {\em
Poissonian} for such $\g$.

Alternatively, if $\g$ consists of a {\em decorated loop}, i.e. a
single loop with attached tree decorations, there are additional
contributions to $\ave{n_{\g}^2}$ to leading order, due to
configurations of $\g_2$ where the two copies of $\g$ share the loop
but not all of the decorations; as a result the asymptotic
distribution of $n_{\g}$ fails to be Poissonian, and is typically
wider.

\subsubsection{Count correlations}

In a similar way, the {\em correlation} between the counts for two
distinct small graphs, $\g$ and $\g'$, say, can be analyzed by
considering the expected value of the product of their counts,
$\ave{n_{\g} n_{\g'}}$.  Again, this can be seen as a sum over their
relative embedding positions and over the absolute position of the
combined graph.

If both graphs are {\em trees}, the leading contribution to
$\ave{n_{\g} n_{\g'}}$ comes from cases where the two subgraphs are
completely disjoint.

In the {\em mixed case} of one graph being a tree, the other a
connected one-loop graph, the leading contribution to $\ave{n_{\g}
n_{\g'}}$ again comes entirely from the completely disjoint case. The
argument can be generalized to higher moments, indicating the
asymptotic lack of correlations between the two counts.

The final case of interest is when {\em both} graphs are connected
one-loop graphs.  If their loops {\em differ} in length, the leading
contribution to $\ave{n_{\g} n_{\g'}}$ again stems entirely from the
completely disjoint cases, and the counts are asymptotically
uncorrelated. If the loops have the {\em same} length, however, there
are additional contributions from cases where the overlap graph
contains the loop, yielding a positive correlation between the two
counts.

For a discussion of subgraph counts in the context of the (not
necessarily sparse) classic model, based on the concepts of balanced
and strictly balanced subgraphs, see e.g. chpt. 4 of
ref. \cite{Boll01}.

\subsection{Subgraphs statistics II: CDRG-s}

Next, we wish to study the statistics of small subgraph counts in a
{\em CDRG-s} ensemble, obtained as the restriction to simple graphs of
the corresponding multigraph ensemble, where simple means the absence
of loops of length one and two.

Thus, we are led to study the distribution of such loops in the
multigraph ensemble, as represented by the subgraph counts when $\g$
is a {\em pure 1-cycle} (vertex with a tadpole) or a {\em pure
2-cycle} (two vertices connected by a double edge).
The relevant results from the previous subsection as applied to these
counts imply the following for a random graph $\G$ from a CDRG-m
ensemble:
\begin{itemize}
\item The expected number $\ave{n_1}$ of 1-cycles in $\G$ is
asymptotically given by $\alpha \equiv \tr\(\T\E\)/2$, and the count
asymptotically follows a Poissonian distribution, $\text{Prob}\(n_1\)
= e^{-\alpha}\alpha^{n_1}/n_1!$.
\item The expected number $\ave{n_2}$ of 2-cycles in $\G$ is
asymptotically given by $\beta \equiv \tr\(\T\E\)^2/4$, and the count
asymptotically follows a Poissonian distribution, $\text{Prob}\(n_2\)
= e^{-\beta}\beta^{n_2}/n_2!$.
\item The 1-cycle and 2-cycle counts are asymptotically uncorrelated,
with each other as well as with the count of any small {\em simple}
graph $\g$ in the form of a tree or a one-loop connected graph.
\end{itemize}
There are two important implications for the corresponding CDRG-s
ensemble:
\begin{itemize}
\item The probability that a random graph from the associated
multigraph ensemble be simple is asymptotically given by
\beq
	\text{Prob(simple)} = e^{-\alpha-\beta},
\eeq
as claimed in ref. \cite{CDRG} without a detailed proof.
\item The count distribution for a {\em simple} small subgraph $\g$ in
a random graph drawn from a CDRG-s ensemble is to leading order {\em
asymptotically identical} to the corresponding distribution in a
random graph drawn from the corresponding CDRG-m ensemble.
\end{itemize}
As a result, the computational rules for subgraph counts given in the
previous subsection apply without modification also to CDRG-s, for the
asymptotically expected subgraph counts of small {\em simple} graphs
to leading order. For a {\em nonsimple} $\g$, the count will of course
vanish identically -- a simple graph has only simple subgraphs.

\section{Global properties}
\label{Global}

The original CDRG article \cite{CDRG} contained a brief generating
function analysis of the asymptotic size distribution of connected
components (clusters). Here we give a more detailed derivation,
combined with a more elaborate analysis of the result.

\subsection{Connected component statistics}

Consider a large random graph $\G$ drawn from an arbitrary CDRG-m
ensemble. Let $P_n$ be the distribution of the number of vertices $0
\le n < \infty$ of a cluster as revealed by starting from a random
vertex in $\G$ and recursively revealing neighbors of previously
revealed vertices. Let $g(z)$ be its generating function,
\beq
\label{g}
	g(z) = \sum_n P_n z^n.
\eeq
At any finite stage in the revelation process, loops in the subgraph
revealed so far are suppressed with factors of $1/N$. Thus, in the
thermodynamic limit we expect the revealed subgraph, as long as it is
finite, to form a tree, and as a result, the following analysis can be
expected to apply equally well to the corresponding ensemble of {\em
simple graphs}.

In terms of the generating function $H(\x)$ for $\{p_{\m}\}$, as
defined in \eq{H}, $g(z)$ can be expressed as
\beq
\label{gh}
	g(z) = z H(\h(z))
\eeq
in terms of the set of similarly defined generating functions $h_a(z)$
for the number of vertices in the subtree revealed by following the
edge emanating from a random stub of given color $a$.
The rationale behind \eq{gh} is that the initial vertex has a random
colored degree $\m$ drawn from the distribution $p_{\m}$. This yields
a factor $z$ for the initial vertex and a factor $h_a(z)$ for each of
its $m_a$ stubs of color $a$; summing the result over $\m$, weighted
with $p_{\m}$, yields \eq{gh}.

The edge functions $h_a(z)$ must satisfy the recursive equations,
\beq
\label{hh}
	h_a(z) = z \sum_b T_{ab} \partial_b H(\h(z)),
\eeq
following from a similar argument: An edge emanating from a stub of
color $a$ has the color $b$ in the other end with probability $T_{ab}
\ave{m_b}$, and is then attached to a vertex with colored degree $\m$
with probability $p_{\m} m_b / \ave{m_b}$. This yields a total factor
of $T_{ab} m_b p_{\m}$. Throw in a factor $z$ to account for that
vertex, and a factor $h_c(z)$ for each subtree reached via one of its
remaining $m_c-\delta_{cb}$ stubs of color $c$; finally, summing over
$b$ and $\m$ yields \eq{hh}.

\subsection{The phase transition and the emergence of the giant}
\label{Giant}

Of particular interest is the result for $z=1$. The recurrence \eq{hh}
for $\h(z)$ for the case of $z=1$ possesses a trivial fixed point
$\h(1)=\1$, yielding $g(1)=1$, expressing the conservation of
probability. However, this fixed point represents the physical
solution only if it is stable, as determined by the Jacobian $\J$
associated with the linearized recurrence in the neighborhood of the
fixed point. $\J$ has the components
\beq
	J_{ab} = \sum_c T_{ac} E_{cb} \; \Leftrightarrow \; \J=\T\E
\eeq
in terms of the matrix $\E$ of second order combinatorial moments.
If all eigenvalues of $\J=\T\E$ are less than unity, the trivial fixed
point $\h(1)=\1$ is stable, and the revelation asymptotically
corresponds to a subcritical branching process, always yielding finite
trees.

Otherwise, the trivial fixed point $\h(1)=\1$ is unstable and will
repel the iterates of the recursion, \eq{hh}. This signals that the
asymptotic branching process is supercritical, with a finite
probability of producing infinite trees. In such a case, a non-trivial
fixed point will appear and attract the iterates, yielding a solution
with $\h_a(1) < 1$, implying $g(1) < 1$ by virtue of \eq{gh}. The
corresponding probability deficit $1-g(1)$ is interpreted as being due
to the existence of a {\em giant component}, and measures the finite
probability that the randomly chosen vertex belongs to the giant,
asymptotically containing a fraction $1-g(1)$ of the vertices.

In analogy to the case of a single color, i.e. DRG, the transition is
typically second order, being due to an initially unstable, nontrivial
fixed point passing the trivial one while they exchange stability
characters -- a transcritical bifurcation in the language of dynamical
systems.

\subsection{Duality}
\label{Duality}

For a {\em supercritical} model, the solution for $g(z)$ resulting
from \eq{gh} for the stable fixed point of the recursion, \eq{hh},
corresponds to a generating function for the contributions from finite
clusters only, and can be shown to emulate another, subcritical CDRG
model -- the {\em dual} model -- as follows.
Define properly normalized functions $\hat{g}(z), \hat{h}_a(z)$ in
terms of the stable solutions $g(z), \h(z)$ as
\begin{subequations}
\beqa
	\hat{h}_a(z) &=& \frac{h_a(z)}{h_a(1)},
\\
	\hat{g}(z) &=& \frac{g(z)}{g(1)}.
\eeqa
\end{subequations}
These will then satisfy (by rewriting (\ref{gh},\ref{hh}))
\begin{subequations}
\beqa
	\hat{g}(z) &=& z \hat{H}(\hat{\h}(z)),
\\
	\hat{h}_a(z) &=& z \sum_b \hat{T}_{ab} \partial_b \hat{H}(\hat{\h}(z)),
\eeqa
\end{subequations}
where $\hat{H}(\x)$ and $\hat{\T}$ are given by
\begin{subequations}
\beqa
	\hat{H}(\x) &=& \frac{H(\h(1)\times\x)}{H(\h(1))},
\\
	\hat{T}_{ab} &=& \frac{1}{h_a(1)} T_{ab} \frac{1}{h_b(1)}.
\eeqa
\end{subequations}
They describe the {\em dual} CDRG model, that is subcritical by
definition: The stable fixed point is mapped to $\hat{\h}(1)=\1
\Rightarrow \hat{g}(1)=1$.  The corresponding transformed CDD is
obtained from the original one by a geometric transform, $\hat{p}_{\m}
\propto p_{\m} \h(1)^{\m}$.  This duality has analogues in other
sparse models, such as DRG (trivially), IRG \cite{Sod02}, and the
classic model \cite{Boll01}.

\section{Redundancy}
\label{Redundancy}

A CDRG model defines a unique ensemble of graphs. The opposite is not
generally true -- there is a built-in redundancy in the CDRG
description, such that several models may describe one and the same
graph ensemble, as we will now demonstrate, based on the local as well
as the global properties of the graphs in a CDRG ensemble, as analyzed
in sections \ref{Local} and \ref{Global} above.

Consider a given asymptotic CDRG model, and define a transformed model
by using a stochastic matrix $\U$, $\U\1=\1$, to define transformed
$H$ and $\T$ as
\begin{subequations}
\beqa
\label{UH}
	\hat{H}(\x) &=& H(\U\x),
\\
\label{UT}
	\hat{\T} &=& \U^{-1}\T\U^{-\top}.
\eeqa
\end{subequations}
This transform conserves the EDD normalization, $H(\1)=1$, and leaves
form-invariant the constraint on $\T$, \eq{Tconstr}.

It also leaves invariant the recursive relations, \eq{hh}, for the
generating functions $\h(z)$ for the size of a subtree found by
following an edge starting from a stub of definite color, if $\h(z)$
is transformed to $\hat{\h}(z) = \U^{-1}\h(z)$. This leaves $g(z)$
invariant by virtue of \eq{gh}, and thus will not affect the
observable distribution of component sizes, $\{P_n\}$.

As for the local properties in the form of expected small subgraph
counts, also these are left invariant, since the computational
(Feynman) rules given in Section \ref{Feynman} invariably yield
expressions in the form of {\em contractions} between the color
indices of combinatorial moments $\E_{(n)}$ on the one hand, and those
of the color preference matrix $\T$ on the other.

The transform can be interpreted as a {\em change of basis} in color
space, such that $U_{a\hat{b}}$ gives the probability $P_{\hat{b}|a}$
that the original color $a$ corresponds to the transformed color
$\hat{b}$.

This suggests the existence of a continuous symmetry group, $\sim
\text{SL}(K-1)$, for the class of CDRG models.  Of course, we have to
be careful to stay in the physical regime, with non-negative values
for $\{T_{ab}\}$ and $\{p_{\m}\}$ (but not necessarily for
$\{U_{ab}\}$ themselves), which restricts the possible transforms and
prevents the class of transformations to form a group. Nevertheless,
it implies that CDRG consists in {\em equivalence classes} of models,
related by transformations of the type (\ref{UH},\ref{UT}).

One can consider even more general transformations, where also the
number of colors, $K$, is changed, requiring a non-square $\U$.  This
enables the reducibility under certain conditions of a model to an
equivalent model with a smaller color space.

\section{Subclasses equivalent to other models}
\label{Relations}

\subsection{DRG}

The restriction of CDRG to a single color, $K=1$, trivially yields
DRG, where a plain degree distribution $\{p_m\}$ is given, while $\T$
reduces to a number $T$, constrained to equal $\ave{m}^{-1}$ by virtue
of \eq{Tconstr}.

More generally, a DRG model effectively results as soon as $\T$ has
rank one, in which case $\T$ takes the form of a direct product,
forced to equal $\T_{\text{DRG}} = \1 \ave{m}^{-1} \t{\1}$, with all
components equal. This prevents the stub colors from affecting the
stub pairing statistics, resulting in a completely random, unbiased
stub pairing.

\subsection{IRG}

Next we wish to identify the CDRG subclass corresponding to IRG. To
that end, consider the restriction of CDRG to ensembles of simple
graphs with a colored degree distribution given by a mixture of
multivariate Poissonians,
\beq
	p_{\m} = \sum_{i=1}^L r_i \prod_a \exp\(-C_{ia}\) C_{ia}^{m_a}/m_a!
\eeq
equivalent to
\beq
\label{HmultiP}
	H(\x) = \sum_{i=1}^L r_i \exp\(\C_{i}\cdot (\x-\1)\)
\eeq
for some $L\ge 1$, where each term in the sum over $i$ corresponds to
a non-negative weight $r_i$ times a normalized multivariate Poissonian
with colored degree average $\ave{\m}_i = \C_i = \{C_{ia}\}$, with the
weights summing up to unity, $\sum_i r_i = 1$.

In ref. \cite{CDRG} the asymptotic equivalence of such an ensemble to
an associated IRG ensemble was shown, based on an analysis of the
equations (\ref{gh},\ref{hh}) for the cluster size distribution.  IRG
(inhomogeneous random graphs) \cite{Sod02} is defined as a colored
extension of the classic model of simple graphs, where a distinct
ensemble of graphs of size $N$ is defined in terms of colored
vertices, where each vertex is independently assigned a color $i\in [1
\dots L]$ according to an arbitrary but fixed distribution
$\{r_i\}$. Then for every pair of vertices, the corresponding edge is
independently realized with a color-dependent probability given by
$c_{ij}/N$, where $i,j$ are the colors assigned to the vertices.

For such a model, a generating function analysis of the cluster size
distribution can be done, analogous to the one represented by
eqs. (\ref{gh},\ref{hh}). The result \cite{Sod02} is that for IRG, the
generating function for the cluster size distribution, $g(z)$ as
defined in \eq{g}, can be written as a weighted sum
\beq
\label{ggi_irg}
	g(z) = \sum_i r_i g_i(z)
\eeq
where $g_i(z)$ is the generating function for the size distribution,
{\em conditional} on the IRG vertex color $i$ of a randomly chosen
initial vertex.  These satisfy a set of recursive relations amounting
to
\beq
\label{gigj_irg}
	g_i(z) = z \exp\(\sum_j c_{ij} r_j \(g_j(z)-1\)\)
\eeq
As shown in ref. \cite{CDRG}, by defining $g_i(z) = z \exp(\sum_a
C_{ia} (h_a(z)-1))$ and $c_{ij} = \sum_{ab} C_{ia} T_{ab} C_{jb}$, eqs
(\ref{gh},\ref{hh}) can be written in the form of eqs. (\ref{ggi_irg},
\ref{gigj_irg}), showing the asymptotic equivalence from the point of
view of cluster size distributions.

An interesting question then is whether this relation persists when
considering small subgraph counts. In the rules for the computation of
the asymptotically expected count of a small subgraph $\g$, as defined
in section \ref{Local}, each vertex in $\g$ with $n$ stubs is
associated with a factor $N\E_{(n)}$. For a CDD as defined by
\eq{HmultiP}, the combinatorial moment $\E_{(n)}$ simplifies to
$\sum_{i=1}^L r_i \C_i^{\circ n}$, where $\C_i^{\circ n}$ stands for
the outer (tensor) product of $n$ factors of $\C_i$, one for every
stub.
Absorbing these stub factors into the edge factor, $\T/N$, yields a
set of Feynman rules with an independent IRG color $i$ for each
vertex, acquiring a corresponding factor $Nr_i$, and a factor of
$\t{\C_i} \T \C_j/N = c_{ij}/N$ for every edge connecting a pair of
vertices with respective IRG colors $i,j$. The product of vertex and
edge factors should be summed over the IRG colors $i,j,\dots$, and the
result divided by the usual symmetry factor $S_{\g}$.

Indeed, these are the correct rules for the expected simple subgraph
counts in an IRG model, as can be derived using simple arguments; this
confirms the asymptotic equivalence between the two models previously
indicated by the cluster size analysis.

\subsection{Other subclasses}

As a special case of the IRG subclass, a CDRG-s ensemble with a CDD in
the form of a single multivariate Poissonian, as defined by
\eq{HmultiP} with a single term, $H(\x) = \exp(\C\cdot (\x-\1))$, is
asymptotically equivalent to the classic model with the parameter
value $c=\t{\C}\T\C = \t{\C}\cdot\1 = \sum_a C_a$.

Another interesting subclass of CDRG is defined by the restriction to
models with {\em monochrome} vertices \cite{Newm03}, such that for
each vertex all its stubs are forced to have the same color.  It is a
trivial exercise to derive the rules for subgraph counts as well as
the equations for the generating function $g(z)$ for the cluster size
distribution in such an ensemble.

In fact, the monochrome subclass is sufficient for spanning IRG, since
for a given IRG model as defined by $\{c_{ij}\},\{r_i\}$, one can
always find an associated CDRG model with the identical color space by
using a {\em diagonal} matrix, $C_{ia} = C_i \delta_{ia}$ with
$C_i=\sum_j c_{ij} r_j$. This yields an equivalent monochrome CDRG
model defined by $T_{ij} = c_{ij}/(C_iC_j)$ and $H(\x) = \sum_i r_i
\exp\(C_i (x_i - 1)\)$.

\section{Concluding Remarks}
\label{Conclusion}

We have considered and analyzed a recently suggested general class of
ensembles, CDRG, of sparse random graphs, based on a hidden coloring
of stubs.  We have extended the formalism to incorporate ensembles of
multigraphs (CDRG-m), in addition to the originally considered
ensembles of simple graphs (CDRG-s).

A distinct random graph model can be defined asymptotically by
specifying a colored degree distribution $\{p_{\m}\}$, controling the
distribution in the number and colors of the connections of a node,
and a color preference matrix $\T$, governing the relative tendency
for connections between stubs with definite pairs of colors.
Based on such an asymptotic model, an ensemble of simple graphs or
multigraphs of a given size can be defined.

For such models, we have demonstrated the calculability of local as
well as global observable structural properties, important for the
anticipated use of the formalism as a target for model infererence
based on the observed properties of real-world networks.

Local graph characteristics can be represented by the statistics of
small subgraph counts. We have derived a set of simple rules for
calculating the asymptotically expected count of an arbitrary small
graph, and demonstrated the equivalence between the two types of
ensembles (of simple or multigraphs) as far as simple subgraphs counts
are concerned.  We have also discussed the shapes of the count
distributions, and shown that a Poissonian distribution results
asymptotically only for simple cycles.  By comparing the expected
counts in DRG and CDRG of certain simple subgraphs, we have
demonstrated the role of the hidden coloring in enabling a non-trivial
edge correlation structure.

Global properties have been exemplified by the statistics of cluster
sizes, for which we have performed a detailed analysis using
generating function techniques. The analysis shows that an arbitrary
CDRG model displays a percolation threshold at a well-defined critical
hypersurface in parameter space, above which a giant component appears
containing a finite fraction of the vertices in the thermodynamic
limit. We have also demonstrated for a supercritical model the
existence of a dual model -- an associated subcritical model
describing the non-giant part.

The algebraic properties of the equations involved in both the local
and global analysis reveal a redundancy (or symmetry) in CDRG, such
that several superficially distinct models describe the same
observable ensemble of graphs. This redundancy can be seen as being
due to the possibility of a change of basis in the abstract color
space.

The rules for the computation of expected subgraph counts have a form
strongly reminiscent of Feynman rules for perturbative calculations in
statistical field theory, indicating a relationship between CDRG
models and field theories, in analogy to the case for DRG
\cite{BuCoKr01}. Work is in progress to explore such relations, and
the results will be presented in a separate article \cite{CDRG-FT}.

The CDRG class of random graph models is very general, and contains
several previously studied models and classes of models as special
cases. Its structure is also such that it should admit a
straightforward generalization e.g. to models of directed graphs.
While CDRG sofar has been considered only for degree distributions
with exponential fall-off for large degrees, it should be extendable
to power-behaved degree distributions if proper care is taken. The key
obstacle (inherited from DRG) is that in such a case the higher
moments of the (colored) degree distribution diverge, which makes some
observables -- in particular for CDRG-s -- very sensitive to the
precise definition of the ensembles.

Anticipating that the formalism can be extended as indicated above, a
few fundamental questions remain to be answered.
{\bf (1)} Is the resulting class ``{\em complete}'', i.e. does it span
 every reasonable model of sparse, truly random graphs? If not, how
 generalize it?
{\bf (2)} Is it unnecessarily general, i.e. can an arbitrary CDRG
 model be reformulated in a simple way entirely in terms of observable
 graph properties, without utilizing hidden variables such as color?

\acknowledgments{An informative discussion with K. Nowicki on the
 complications associated with subgraph distributions is gratefully
 acknowledged. This work was in part supported by the Swedish
 Foundation for Strategic Research.}


\end{document}